\documentclass[10pt,twocolumn]{IEEEtran}
\usepackage[T1]{fontenc}% optional T1 font encoding
\usepackage{amsthm}

\ifCLASSINFOpdf
\else
\fi

\usepackage{cite}
\usepackage[cmex10]{amsmath}
\usepackage{wasysym}
\usepackage{amsthm}
\usepackage{setspace}
\usepackage{amssymb}
\usepackage{algorithmic}
\usepackage{algorithm}
\usepackage{enumerate}
\usepackage{stfloats}
\usepackage{multicol,multienum}
\usepackage[dvips]{graphicx}
\usepackage{booktabs}
\usepackage[dvips]{epsfig}
\usepackage{subfigure}
\usepackage{multirow}
\usepackage{float}
\usepackage{xcolor}
\usepackage{bm}
\usepackage{array}
\usepackage{multirow}
\usepackage{makecell}
\usepackage{array}

\hfuzz=\maxdimen
\begin{document}
%\title{A Dynamic User-Scheduling-based Hierarchical Power Control in Two-tier Femtocell Networks from the Perspective of Energy Efficiency}
\title{Game-theoretic Learning Anti-jamming Approaches in Wireless Networks}

\author{
       Luliang~Jia,
       Nan~Qi,
       Feihuang~Chu,
        Shengliang~Fang,
     %   Yuhua~Xu,
        Ximing~Wang,
        Shuli~Ma,
        and~Shuo~Feng
}

\maketitle


\vspace{-0.8cm}

\begin{abstract}
In this article, the anti-jamming communication problem is investigated from a game-theoretic learning perspective. By exploring and analyzing intelligent anti-jamming communication, we present the characteristics of jammers and the requirements of an intelligent anti-jamming approach. Such approach is required of self-sensing, self-decision making, self-coordination, self-evaluation, and learning ability. Then, a game-theoretic learning anti-jamming (GTLAJ) paradigm is proposed, and its framework and challenges of GTLAJ are introduced. Moreover, through three cases, i.e., Stackelberg anti-jamming game, Markov anti-jamming game and hypergraph-based anti-jamming game, different anti-jamming game models and applications are discussed, and some future directions are presented.
\end{abstract}

% \fi
\IEEEpeerreviewmaketitle

\vspace{-0.5cm}
\section{Introduction}
\vspace{-0.1cm}

Due to the sharing and open nature, wireless communications are vulnerable to jamming attack, which has been a critical and major threat that captured growing attention in the past decade [1]-[4]. Various anti-jamming approaches have been proposed to cope with jamming attacks, which can be divided into five categories: confrontation, avoidance, elimination, hide, and deceit [5]. Many common anti-jamming countermeasures are adopted, such as frequency hopping spread spectrum (FHSS) and direct sequence spread spectrum (DSSS). However, traditional anti-jamming approaches are in low spectral  efficiency, and sacrifice resource utilization efficiency for transmission reliability. Moreover, they employ fixed or preset patterns that lack of intelligent decision-making ability. Thus, it is of great importance to develop effective anti-jamming approaches to fight against jamming attacks, especially for scenarios with limited resources and intelligent jammers.

With the rapid development of cognitive radio and artificial intelligence technology, the anti-jamming problem has been endowed with new content. On the one hand, jamming attacks can utilize these emerging technologies to produce more complex and effective jamming patterns. On the other hand, these new technologies also provide new ideas to improve the anti-jamming ability for wireless communication systems. Thus, it is a trend to learn from cognitive radio and artificial intelligence technologies. In particular, the game-theoretic learning method is an extremely promising technology, since it has the following advantages in anti-jamming field:

\begin{itemize}
\item  	A jammer aims to intentionally disrupt the transmission of legitimate users, making their communication ability reduced or lost. However, legitimate users pursue eliminating the impacts of various jamming attacks as much as possible to achieve reliable information transmission. Due to the natural confrontation characteristic, there are interactions between legitimate users and jammers. Game theory stands out as a well-developed tool to formulate the mutual interactions among players, and it can adequately formulate and analyze interactions between legitimate users and jammers [4]-[12].
\item  	It is difficult to obtain accurate information of the opponents for both legitimate users and jammers. Moreover, the jamming environment is dynamic because of the jamming activities and time-varying electromagnetic environment. Thus, the anti-jamming problem needs to deal with dynamic and incomplete (DI) information constraints. Learning is an effective method that can cope with the DI information constraints through trial and error interaction with the jamming environment \cite{existwork9}.
\end{itemize}

 Game theory has provided insights to improve anti-jamming performance. In \cite{existwork6}, an anti-jamming game was formulated, and  a coordination-learning approach was proposed. In \cite{existwork7}, a bimatrix game was employed to analyze and model the interaction between the legitimate link and the jammer. In \cite{existwork8}, a Stackelberg power game with observation errors was investigated. In \cite{existwork9}-\cite{existwork12}, the authors investigated some preliminary exploration on game-theoretic learning anti-jamming approach. These existing works are summarized in Table I.

\begin{table*}[htbp]
	\centering
	\caption{Summary of game theory-based anti-jamming approaches.}
	\resizebox{\textwidth}{23mm}{
	\begin{tabular}{|p{0.03\textwidth}|p{0.13\textwidth}|p{0.18\textwidth}|p{0.15\textwidth}|p{0.33\textwidth}|}
		\toprule
		Ref & Models & Techniques & Situations & Advantages \\
		\midrule
		\cite{existwork6} & Correlated equilibrium game & Coordination-learning & Interference and jamming & Taking jamming signals as coordination signals.   \\
\hline
		\cite{existwork7}  & Bimatrix game & Linear programs & Jamming & Describing confrontation relationship.  \\
\hline
	    \cite{existwork8}, \cite{existwork9}  & Stackelberg game &Convex optimization; Hierarchical learning & Jamming [8];Interference and jamming [9] & Capturing hierarchical behaviors; Analyzing competitive interactions at different levels. \\
\hline
		\cite{existwork10}  & Hypergraph game & Stochastic learning & Interference and jamming & Capturing accurate interference relationship.  \\
\hline
		\cite{existwork11}  & Markov game & Multi-agent Q-learning & Interference and jamming & Describing the dynamic characteristic of jamming environment.  \\
		\bottomrule
	\end{tabular}}
\end{table*}

Motivated by the above, we propose a new anti-jamming paradigm called game-theoretic learning anti-jamming (GTLAJ) communication, which integrates game theory into solving the anti-jamming problem. Note that the GTLAJ is a new paradigm to design effective anti-jamming communication approaches. The main contributions of this article are:
\begin{itemize}
\item A framework of intelligent anti-jamming communication is introduced. Then, characteristics of jammers and requirements of such intelligent anti-jamming communication approach are analyzed and outlined.
\item A game-theoretic learning anti-jamming paradigm is proposed, and different anti-jamming games are analyzed.
\item Three cases, i.e., Stackelberg anti-jamming game, Markov anti-jamming game, and hypergraph-based anti-jamming game, are given, and future directions are illustrated.
\end{itemize}

%The rest of the article is organized as follows. In Section II, we discuss and analyze the technical challenges and fundamental requirements of anti-jamming defence in wireless networks. In Section III, the Stackelberg game-theoretic model is investigated, and an anti-jamming decision-making framework based on Stackelberg game is established. In Section IV, two preliminary case studies are given, and future research directions are discussed in Section V. Finally, concluding remarks are presented in Section VI.

%\vspace{-0.4cm}
\section{Requirements and Summary of Intelligent Anti-jamming Communication}

To achieve reliable transmission in the increasingly complex and harsh electromagnetic environment, the intelligent anti-jamming communication technology is promising. A compositional architecture of an intelligent anti-jamming communication system is shown in Fig.1. Similar to a cognitive dynamic system \cite{existwork14}, an intelligent system should have five distinctive characteristics: perception-action cycle (PAC), memory, attention, intelligence, and language. The PAC begins with the observation of the radio environment. Then, legitimate user acts according to the feedback information. Given a radio environment, memory means storage and updating of acquired knowledge, which is essential to learn from the jamming environment. Attention is another mechanism, which determines the priority of available resources allocation. Different priorities indicate different degrees of importance, and high priority users need to be satisfied first. Intelligence is the most important property that based on PAC, attention, and memory. According to a large amount of feedback information, it promotes the intelligence level of the system, and then makes intelligent decision. Similar to the role in human brain, language is an important tool to connect different parts of a single intelligent system or among multiple intelligent systems.

Inspired by the PAC in cognition cycle, in this article, we focus on the PAC in anti-jamming communication cycle, which plays a significant role in an intelligent anti-jamming system. PAC observes and evaluates the jamming environment, and acts according to the collected perception information.

\begin{figure*}[!t]
\centering
\includegraphics[width=0.9\linewidth]{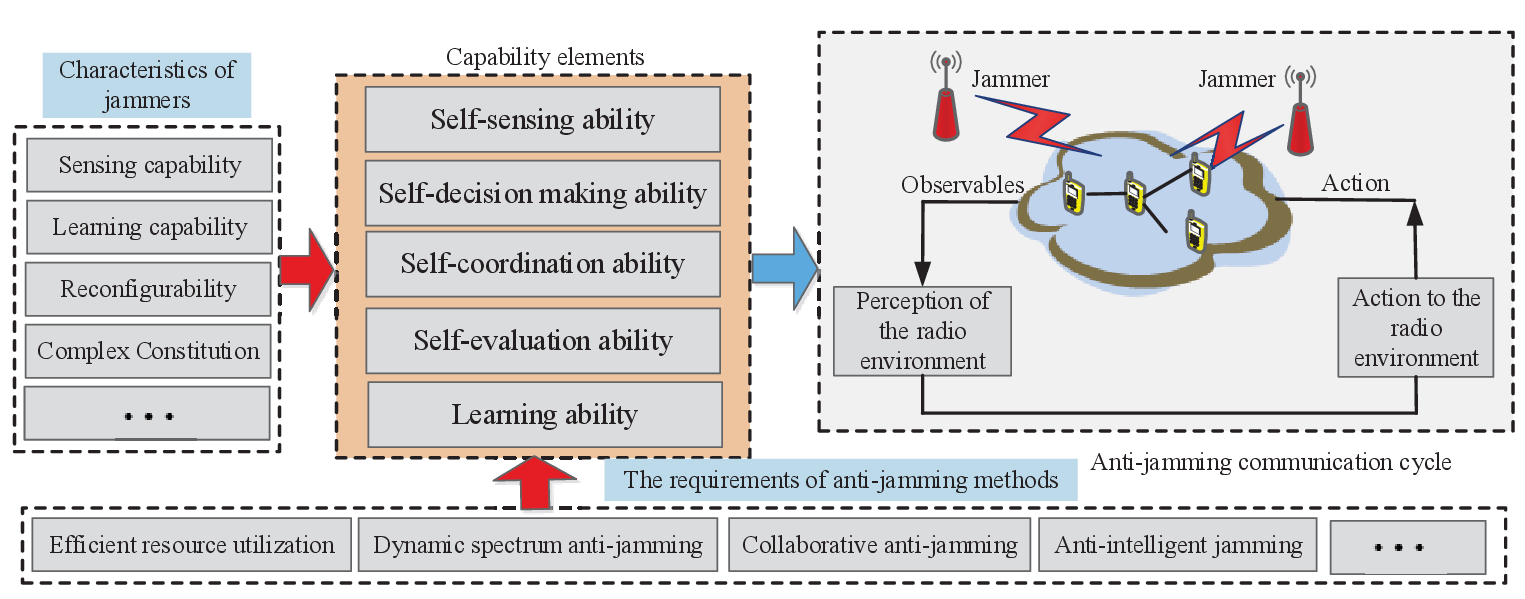}
\caption{The compositional architecture of an intelligent anti-jamming communication system.}
\label{Fig1}
\end{figure*}

%\vspace{-0.5cm}
\subsection{Discussion of the characteristics of jammers}
%\vspace{-0.1cm}
%%%%%%%%%%%%%%%%%%%%%%%%%%%%%%%%%%%%%%%%%%%%%%%%%%%%%%
With the improvement of jamming intelligence level, wireless communication is facing increasingly complex and harsh electromagnetic environment. According to the analysis in [4], an intelligent jammer is of sensing ability, learning ability, and reconfigurability. It should be noted that ordinary jamming pattern can be regarded as a special case of intelligent jammer.

By sensing, it observes or measures the radio environment, and obtains historical knowledge. Based on sensing results, the learning ability means that the jammer mines the communication rule to release more effective jamming patterns. The reconfigurability represents that the jammer can reconfigure its jamming pattern, which can make it more flexible. Moreover, with increasing demands of wireless services, the number of wireless devices increases rapidly. Accordingly, the spectrum resources become very rare, and the corresponding inter-user mutual interference becomes serious. The anti-jamming communication is faced with intense external confrontation and severe internal conflict. Therefore, there is a complex constitution that inter-user mutual interference and external malicious jamming will coexist in the new jamming environment. In addition, the mutual interference problem will become more prominent in dense wireless networks. In this article, the mutual interference refers to co-channel interference.

%\vspace{-0.4cm}
\subsection{The requirements of intelligent anti-jamming approaches}
%\vspace{-0.1cm}
Generally speaking, the development of anti-jamming communication technologies has gone through three stages: fixed spectrum anti-jamming, adaptive spectrum anti-jamming, and intelligent anti-jamming. The fixed spectrum anti-jamming approach means that the spectrum resources do not change according to the external jamming environment, but depends on the anti-jamming ability of the technology itself, such as DSSS and FHSS. As a contrast, adaptive anti-jamming approach is able to sense the external jamming environment and adjust some communication parameters in quasi-real time according to the jamming activities, such as adaptive FHSS, and adaptive DSSS. Intelligent anti-jamming is an inevitable trend of anti-jamming technology. It can recognize and learn the external jamming environment, and make decisions autonomously. For an intelligent anti-jamming system, it should meet the following requirements.

\textbf{Effective resource utilization:} The spread-spectrum-based anti-jamming technologies are common, and widely used in military and civil communication systems. They improve the anti-jamming performance at the cost of spectrum resource utilization efficiency, making the contradiction between anti-jamming performance and spectrum resource scarcity increasingly acute. It is an important criterion of anti-jamming design to obtain better anti-jamming performance with minimum resource cost. Thus, it is necessary to develop optimal resource allocation based anti-jamming schemes to cope with the complicated electromagnetic environment, as such higher resource utilization efficiency can be expected.
%The reason is that spectrum resource is limited, and it becomes more scarce with the increase of frequency equipments

\textbf{Dynamic spectrum anti-jamming:} To deal with the dynamic characteristic of the jamming environment, it is necessary to consider the dynamic mechanism and weakness of jammers. Based on the idea of dynamic spectrum access in cognitive radio, the dynamic strategy is adopted to resist and avoid jamming attacks. Different from traditional fixed or preset anti-jamming patterns, it aims to flexibly utilize spectrum resources, and can adapt to varying spectrum environment. Moreover, it is more difficult for jamming attacks to master the communication rule.

\textbf{Collaborative anti-jamming:} To enhance the anti-jamming performance, it is necessary to deeply explore a multi-level collaborative mode and design a collaborative anti-jamming paradigm, such as information exchange, local interaction and jamming utilization. Based on collaboration mechanism, it can realize the transformation of anti-jamming ideas from independent countermeasures to collaborative countermeasures. In \cite{existwork11}, the information exchange among users is adopted. In \cite{existwork10}, local interaction is employed, and each user considers the impact on neighbor users. In \cite{existwork6}, jamming signals are regarded as coordination signals to guide anti-jamming spectrum access.

%it is necessary to design anti-jamming methods based on internal mutual interference collaboration and external malicious jamming confrontation. Based on collaboration among users, it can effectively deal with the problem of interference elimination. There are different understandings of collaboration in anti-jamming field, such as, and so on.

\textbf{Anti-intelligent jamming:} With the high integration of artificial intelligence technology and communication jamming technology, intelligent jamming attacks become serious threats that deteriorate information transmission. As jamming environment becomes more dynamic, and the jamming rule is more difficult to mine. Anti-intelligent jamming problem becomes important in anti-jamming field, and will be an inevitable trend of intelligent anti-intelligent jamming attacks.

\vspace{-0.3cm}
\subsection{The capability elements of intelligent anti-jamming approaches}

Based on the PAC, an intelligent anti-jamming system should be equipped with the following abilities.

\textbf{Self-sensing ability:} It is the base for an intelligent anti-jamming system that aims to autonomously observe the jamming environment and detect jammers' activities.

\textbf{Self-decision making ability:} It is the core of an intelligent anti-jamming system. Based on sensing results, it autonomously makes decision and obtains the desirable anti-jamming strategies.

\textbf{Self-coordination ability:} With the increasing of wireless devices, mutual interference becomes a serious problem. Thus, it is necessary to coordinate different users for interference mitigation. Moreover, the multi-level collaboration mode, i.e., information exchange, local interaction, and relay enhancement, can be employed to improve the anti-jamming ability.

\textbf{Self-evaluation ability:} To estimate the effectiveness of anti-jamming approaches, anti-jamming evaluation is indispensable. Through online real-time evaluation, the effectiveness of anti-jamming approaches can be measured, and anti-jamming schemes can be adjusted according to the evaluation in real-time to further improve the anti-jamming performance.

\textbf{Learning ability:} Learning is the critical ability of an intelligent anti-jamming system that aims to acquire experiences, mines jamming rules, and accumulates historical knowledge. The higher goal is to predict jamming rules and discover jamming knowledge according to the historical information.

%\vspace{-0.3cm}
\section{Game-theoretic learning anti-jamming communications: framework and challenges}

\subsection{Discussion of framework}
 A framework of GTLAJ is designed and illustrated in Fig. 2, and it mainly includes two parts: anti-jamming game formulation, and design of anti-jamming learning algorithms.

\begin{figure}[!t]
\centering
\includegraphics[width=1.0\linewidth]{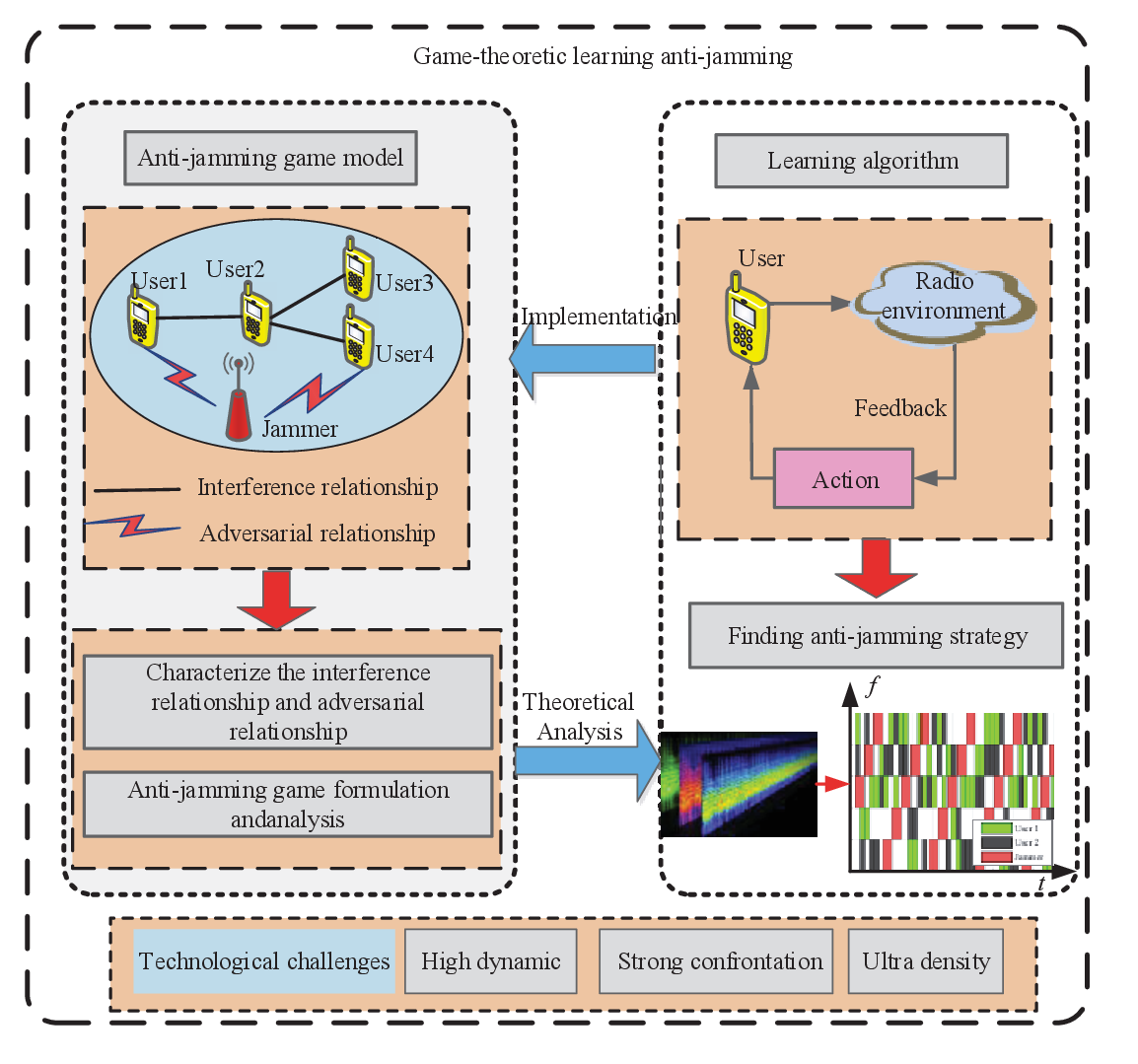}
\caption{Framework of game-theoretic learning anti-jamming communication.}
\label{Fig2}
\end{figure}

\textbf{Anti-jamming game formulation:} The anti-jamming game model focuses on theoretical analysis and provides a theoretical framework for analyzing anti-jamming problem. In the anti-jamming field, many game models can be employed, such as Stackelberg game, graphical game, and Markov game. A suitable game model can be selected according to specific anti-jamming scenarios: Stackelberg game can model the sequential interactions between legitimate users and jammers; Markov game is a suitable tool to characterize the dynamic activities of jammers; the graphical/hypergraphical game can accurately model the mutual interference among legitimate users.

According to whether the jammer is a game player or not, anti-jamming game models can be divided into two types.

\textbf{Jammer as a player:} Stackelberg game is a suitable tool to analyze and model the anti-jamming problem when the jammer acts as a player \cite{existwork9}. A Stackelberg anti-jamming game is an ideal game framework that includes two types of players with different attributes, namely, jammers and legitimate users. They are both game players that aim to optimize their own utility functions. The game can capture the sequential interactions between legitimate users and jammers. Moreover, it can not only characterize the adversarial relationship between legitimate users and jammers, but also describe the competitive behavior among legitimate users.

\textbf{Jammer as environment:}
From the perspective of engineering implementation, it is difficult to obtain accurate jammer information in some scenarios, in which jammers are often regarded as part of the environment. As an important game model, Markov game stands out to formulate the anti-jamming problem that can describe the dynamic characteristics of jammers in \cite{existwork11}. There are also other game models that can be employed in the anti-jamming problem. For example, in \cite{existwork15}, graphical game can be adopted to characterize the mutual interference among users. Furthermore, to represent the mutual interference more accurately, hypergraph game is adopted, which can simultaneously capture both strong interference relationship and cumulative weak interference relationship in \cite{existwork10}, especially in ultra-dense networks.

In an anti-jamming game model, it first needs to characterize two kinds of relationship: one is the adversarial relationship between jammers and legitimate users, and the other is the mutual interference relationship among users. Second, it is necessary to formulate an anti-jamming game that is generally non-cooperative. Game models pursue steady solutions, while people usually want to get the optimal solutions in practical scenarios. Unfortunately, steady solutions are usually not the optimal ones. Therefore, the first and core problem is how to make these steady solutions equal or close to the optimal ones in the anti-jamming game formulation. It should be noted that the properties of anti-jamming game models are closely related to the utility function, which determines the performance of game models.

\textbf{Design of anti-jamming learning algorithms:} Learning algorithms are dedicated to designing effective anti-jamming algorithms, and can converge to a stable solution (i.e., Nash equilibrium, \emph{NE}) with good performance. To acquire solutions of an anti-jamming game, learning algorithms need to be designed to attain desirable anti-jamming strategies through some behavior updating rules [13]. Some learning algorithms, such as spatial adaptive play, multi-agent learning, and stochastic learning automata, have been presented in existing works. It should be noted that proving the validity and existence of game solutions is important, though direct proof is difficult. The usual way is to prove the formulated anti-jamming game as a specific type of game, and then guarantee the existence and effectiveness of the solution according to the properties of the specific game model. For example, when the formulated game is an exact potential game, it means that it has at least one pure strategy NE, and global or local optimal solution of the potential function is a pure strategy NE. Therefore, the important problem is how to design the potential function so that the objective function of the problem is linearly related to the potential function.

\vspace{-0.4cm}
\subsection{Discussion of technical challenges}

Although GTLAJ has many advantages, there are still some technical challenges. In this section, we summarize the main challenges from the perspective of network characteristic.

\textbf{High dynamic:} Due to the time-varying electromagnetic spectrum environment, being dynamic is its inherent attribute in wireless networks, and the channel state may change from time to time. Moreover, the dynamic can be greatly significant considering the dynamic characteristics of jammers, such as jamming channel, jamming power, and jamming pattern. In addition, the dynamic characteristics of the wireless network itself should be considered, such as traffic demands and network topology relationship.

\textbf{Strong confrontation:} Confrontation is the most important feature due to the adversarial and non-cooperative relationship between legitimate users and jammers. Legitimate users aim to achieve reliable transmission, while jammers aim to disrupt such communication. Besides, such confrontation becomes more intense as the intelligent level of jammers improves.

\textbf{Ultra density:} Various wireless devices are deployed in some wireless networks, and both mutual interference among users and malicious jamming should be considered. The malicious jamming is the most immediate threat that deteriorates the transmission of legitimate users on the one hand, and on the other hand, the mutual interference is an important factor that restricts the system performance. This will be increasingly prominent in ultra-dense wireless networks, such as small cell networks and unmanned aerial vehicles networks. In addition, it should be noted that there are not only many strong interference relations, but also many weak interference relations in ultra-dense networks.

%\vspace{-0.3cm}
 \section{Case studies}

 \subsection{Stackelberg anti-jamming game}

 Stackelberg game is an extension of the non-cooperative game, and contains two types of game players, the leaders and the followers. The leaders take actions first, and the followers play their strategies according to the leaders' announced strategies.  In a Stackelberg anti-jamming game, it has two types of players, namely, malicious jammers and legitimate users. In this game, both legitimate users and jammers have their own utility functions, and aim to maximize their own utility functions. The Stackelberg anti-jamming game has two advantages in the anti-jamming field. First, it can model and analyze the common hierarchical behaviors between legitimate users and jammers in an anti-jamming problem. For example, legitimate users need to detect the jammers' activities in some scenarios. Moreover, intelligent jammers can learn legitimate users' actions. Second, the Stackelberg game can characterize the competitive interactions at different levels. The competition exists between leader players and follower players, that is, between legitimate users and jammers. Besides, there are also mutual competitions among legitimate users at the same level.

 In \cite{existwork9}, there are \emph{N} legitimate users and one jammer in a system, and a Stackelberg anti-jamming game with one leader and multiple followers is investigated. Specifically, the jammer is the leader, and the legitimate users are followers. Then, to obtain the desirable anti-jamming strategies, a hierarchical learning framework is formulated, and a hierarchical learning algorithm is designed. For comparison, four methods are considered: the proposed algorithm, the best NE, the worst NE, and the random selection method. For the best and the worst NE, the best response algorithm is employed. 200 independent trials are simulated, and then the best one and the worst one are respectively taken. As can be known in Fig. 3, the best and the worst NE are respectively the upper and lower bounds of the proposed algorithm. Compared with the random selection approach, the proposed algorithm has a higher expected achievable rate. The reason is that each legitimate user randomly chooses one channel in each time slot for random selection approach, and the proposed algorithm can converge to a desirable anti-jamming strategy. In addition, the expected achievable rate increases with lower jamming power \emph{J}.

%%%%%%%%%%ͨÀ¸Í¼Æ¬ÅÅ°æ%%%%%%
%%%%%%%%%%%%%%%%%%%%%%%%%%%%
%\begin{figure*}[!t]
%\subfigure[A framework of Stackelberg anti-jamming game.]{
%\label{fig:mini:subfig:a} %% label for first subfigure
%\begin{minipage}[b]{0.40\textwidth}
%\centering
%\includegraphics[width=\linewidth]{Fig3a}
%\end{minipage}}%
%\subfigure[Performance comparison of different algorithms.]{
%\label{fig:mini:subfig:b} %% label for second subfigure
%\begin{minipage}[b]{0.41\textwidth}
%\centering
%\includegraphics[width=\linewidth]{Fig3b}
%\end{minipage}}
%\caption{Stackelberg anti-jamming game.}
%\label{performance} %% label for entire figure
%\end{figure*}
%%%%%%%%%%ͨÀ¸Í¼Æ¬ÅÅ°æ½áÊø%%%%%%

\begin{figure}[!t]
\centering
\includegraphics[width=\linewidth]{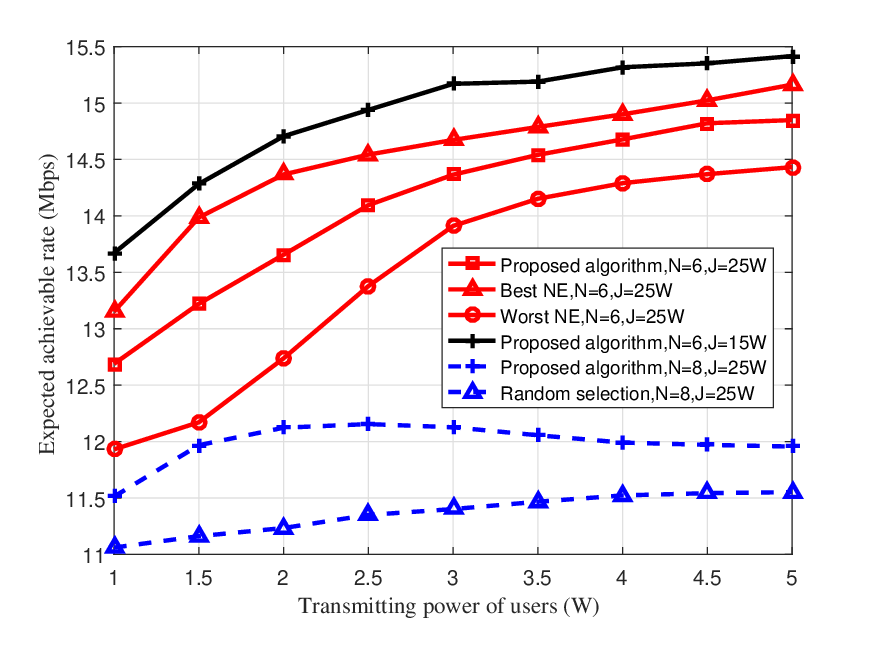}
\caption{Performance comparison of different algorithms in the Stackelberg anti-jamming game.}
\label{Fig3b}
\end{figure}

%\vspace{-0.4cm}
\subsection{Markov anti-jamming game}

 Markov game is an extension of the Markov decision problem with many agents. This game has many advantages. First, the jamming activities are usually dynamic, and the Markov game can describe the dynamic characteristics of a jamming environment, which provides a mathematical framework for decision optimization in dynamic scenarios. Second, Markov game can represent the relationship among players, and describe collaboration and competition among players according to their utility functions. Moreover, the multi-agent reinforcement learning approach is a suitable tool to obtain desirable anti-jamming strategies through "action-feedback-action" approach, in which each legitimate user acts as an agent.

 In \cite{existwork11}, a Markov anti-jamming channel selection game is formulated. Then, a collaborative multi-agent reinforcement learning anti-jamming algorithm is proposed. The coordination among users is considered through information exchange (e.g., Q table), which can be realized by a common control channel. Based on collaborative learning, it can simultaneously tackle the jamming and mutual interference among legitimate users. Fig. 4 shows the performance comparison of the average rate in sweep jamming and comb jamming scenarios. Compared with the sensing-based method, the independent \emph{Q}-learning and the random selection method, the proposed algorithm has the highest average rate in the sweep jamming scenario. The reason is that each legitimate user independently employs a standard \emph{Q}-learning for the independent \emph{Q}-learning method, without considering the coordination; for the sensing-based method, it makes the channel selection depends only on sensing results; in random selection scheme, each legitimate user randomly chooses a channel, and it lacks of learning ability and collaboration mechanism. Moreover, the proposed algorithm also has good performance in the comb jamming scenario.

%%%%%%%%%%ͨÀ¸Í¼Æ¬ÅÅ°æ%%%%%%
%%%%%%%%%%%%%%%%%%%%%%%%%%%%
%\begin{figure*}[!t]
%\subfigure[A framework of Markov anti-jamming game.]{
%\label{fig:mini:subfig:a} %% label for first subfigure
%\begin{minipage}[b]{0.42\textwidth}
%\centering
%\includegraphics[width=\linewidth]{Fig4a}
%\end{minipage}}%
%\subfigure[Performance comparison of different algorithms.]{
%\label{fig:mini:subfig:b} %% label for second subfigure
%\begin{minipage}[b]{0.41\textwidth}
%\centering
%\includegraphics[width=\linewidth]{Fig4b}
%\end{minipage}}
%\caption{Markov anti-jamming game.}
%\label{performance} %% label for entire figure
%\end{figure*}
%%%%%%%%%%ͨÀ¸Í¼Æ¬ÅÅ°æ½áÊø%%%%%%

\begin{figure}[!t]
\centering
\includegraphics[width=\linewidth]{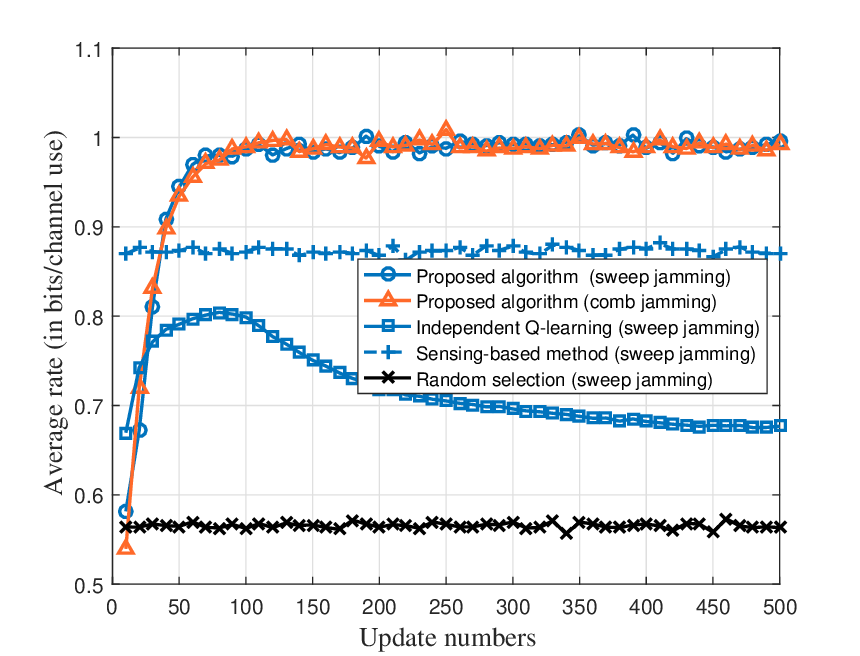}
\caption{Performance comparison of different algorithms in the Markov anti-jamming game.}
\label{Fig4b}
\end{figure}

%\vspace{-0.4cm}
\subsection{Hypergraph-based anti-jamming game}
In ultra-dense wireless networks, anti-jamming defense is more challenging, and mutual interference becomes very prominent. First, the strong interference relationship between two neighboring legitimate users needs to be considered. Second, there is also some accumulative weak interference relationship that may be equivalent to a strong one when the accumulative weak interference from multiple users reaches a certain threshold. However, only the strong interference relationship is considered in traditional graph model. Similar to [10], the strong interference relationship is represented by lines, while the weak accumulative interference relationship is denoted by circles. The lines and circles together form hyperedges. A strong interference relationship occurs between two neighboring legitimate users when they choose the same channel, while a weak accumulative interference relationship is generated among three or more legitimate users. Specifically, only when three or more legitimate users employ the same channel simultaneously, a weak accumulative interference relationship can be generated. Otherwise, there won't be an interference relationship in an accumulative interference relationship hyperedge. For a hypergraph-based anti-jamming game, it has two advantages. First, it can accurately describe the interference relationship among legitimate users, considering both the strong and accumulative weak interference relationships. Second, the anti-jamming problem can be formulated as a hypergraph game, and transformed into a generalized interference and jamming minimization problem.

In \cite{existwork10}, a system with \emph{N} users and \emph{K} jammers is considered, and the number of available channels is \emph{M}. Then, a hypergraph-based anti-jamming game is formulated, and an anti-jamming channel selection algorithm is proposed to achieve the desirable anti-jamming strategies. In the proposed algorithm, each user updates its strategies according to the learning rule based on stochastic learning automata, and a common control channel is needed to realize the information exchange. Fig 5 shows the performance comparison of the expected normalized network capacity. To evaluate the performance of the proposed algorithm, the performance of the graph-based method, and the random selection method are presented for comparison. It shows that the performance of the proposed hypergraph-based algorithm is the best. Besides, the performance improvement is more significant with the increases of hyperedges for the weak accumulative interference (HWI) relationship when the hyperedges of strong interference relationship is fixed. The reason is that the graph-based method ignores the weak accumulative interference relationship, and the random selection scheme is an instinctive method. In addition, growing channels results in the increase of expected normalized network capacity. The reason is that the more the channels, the more available resources for the users, leading to less mutual interference among users. It is also noted that the expected normalized network capacity decrease with lower active probability \emph{p}. The reason is that active probability means a user competes for channels with certain probability, that is to say, a user is active with probability \emph{p}, and the lower active probability brings less user communication.

%%%%%%%%%%ͨÀ¸Í¼Æ¬ÅÅ°æ%%%%%%
%%%%%%%%%%%%%%%%%%%%%%%%%%%%
%\begin{figure*}[!t]
%\subfigure[A framework of hypergraph-based anti-jamming game.]{
%\label{fig:mini:subfig:a} %% label for first subfigure
%\begin{minipage}[b]{0.34\textwidth}
%\centering
%\includegraphics[width=\linewidth]{Fig5a}
%\end{minipage}}%
%\subfigure[Performance comparison of different algorithms.]{
%\label{fig:mini:subfig:b} %% label for second subfigure
%\begin{minipage}[b]{0.56\textwidth}
%\centering
%%\includegraphics[width=\linewidth]{Fig4b}
%%\end{minipage}}
%%\subfigure[Comparison results of the expected achievable rate for different solutions (${P_n} = 4W$).]{
%%\label{fig:mini:subfig:c} %% label for second subfigure
%%\begin{minipage}[b]{0.32\textwidth}
%%\centering
%\includegraphics[width=\linewidth]{Fig5b}
%\end{minipage}}
%\caption{Hypergraph-based anti-jamming game.}
%\label{performance} %% label for entire figure
%\end{figure*}
%%%%%%%%%%ͨÀ¸Í¼Æ¬ÅÅ°æ½áÊø%%%%%%

\begin{figure}[!t]
\centering
\includegraphics[width=\linewidth]{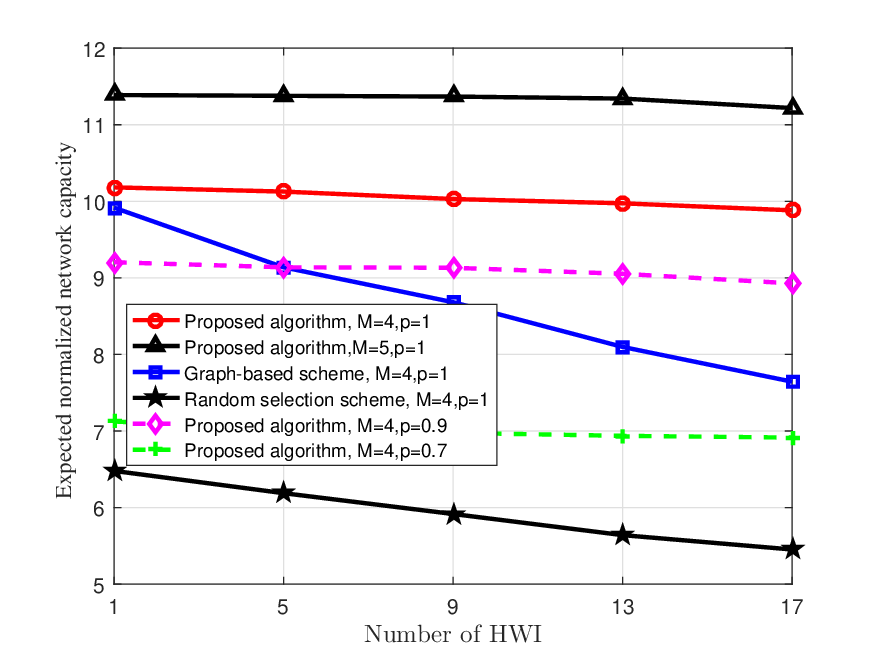}
\caption{Performance comparison of different algorithms in the hypergraph-based anti-jamming game.}
\label{Fig5b}
\end{figure}

 \section{Future research directions}
%%%%%%%%%%%%%%%%%%%%%%%%
Although the GTLAJ approaches attained much attention in the past years, there are still many problems to be discussed.

\begin{itemize}
\item \textbf{Jamming prediction:} Although jammers' activities are dynamic, the jamming spectrum information may have some correlations at different time slots. According to the observed historical information of jammers, useful jamming knowledge and jamming rules can be mined. Then, based on the machine learning method (i.e., long-short-term-memory), the law of jammers' actions can be learned so that jamming behaviors can be avoided in advance. Thus, the improvement from passive anti-jamming to active anti-jamming can be realized.
\item \textbf{Integrated design of jamming attack and anti-jamming:} It is necessary to design the integration of attack and defense. On the one hand, it is essential to ensure communication reliability, and on the other hand, the offensive capability is also required. By utilizing offense capability to suppress opponent transmissions, one user may gain advantages in information transmission by exerting jamming on opponent users. If the opponent user is attacked, it releases the jammed channel due to the friendly jammer, which may be helpful to strive for more channel resources for the users.
\item \textbf{Knowledge transfer:} In an unknown jamming environment, learning technology is important to obtain desirable anti-jamming strategies through continuous trial-and-error interactions. To speed up the convergence speed of learning algorithm, knowledge transfer (i.e., hotbooting technique) is important. When encountering unknown jamming environment, the similarity analysis can be carried out first, and the acquired experiences can be transferred to the new scene. Therefore, it realizes the accumulation of jamming experiences and improves the speed of learning.
\item \textbf{Jamming location:}  Jamming location is helpful to obtain accurate jamming information, and facilitates the anti-jamming method design. To obtain effective anti-jamming strategies, it is necessary to estimate jamming parameters, such as jamming power, and jamming channel gain. The inaccurate estimation is a main challenge in jammer localization due to the confrontation and non-cooperative relationships between legitimate users and jammers. In the process of designing anti-jamming approaches, jamming localization-assisted anti-jamming methods may effectively improve the anti-jamming performance.
\end{itemize}

%\vspace{-0.3cm}
\section{Conclusions}
In this article, the game-theoretic learning anti-jamming (GTLAJ), a new paradigm for intelligent anti-jamming communication, was investigated. First, jammers' characteristics, requirements and capability elements of intelligent anti-jamming approaches were presented. Then, the framework and challenges for GTLAJ were introduced, and different anti-jamming game models were analyzed. Finally, three cases were studied, and future research directions were discussed.

%\vspace{-0.3cm}
\section{Acknowledgements}
This work was supported in part by the National Science Foundation of China under Grant 61901523.

\ifCLASSOPTIONcaptionsoff
  \newpage
\fi

\begin{spacing}{0.9}
\fontsize{8.5pt}{\baselineskip}\selectfont \emph{Luliang Jia} (zz$\_$lljia@nudt.edu.cn) received his M.S. and Ph.D. degree from Army Engineering University of PLA, Nanjing, China, in 2014 and 2018, respectively. Currently, he is an Assistant Professor at Space Engineering University, Beijing, China. His research interests include game theory, learning theory, and communication anti-jamming technology.\\
\end{spacing}

%{Luliang Jia} (jiallts@163.com) received his M.S. degree in Communications and Information Systems from College of Communication Engineering, PLA University of Science and Technology, Nanjing, China, in 2014, and the Ph.D. degree in Communications and Information Systems from College of Communication Engineering, Army Engineering University of PLA. Currently, he is an Assistant Professor at School of Space Information, Space Engineering University, Beijing, China. His research interests include game theory, learning theory, satellite communication, and communication anti-jamming technology.

\begin{spacing}{0.9}
\fontsize{8.5pt}{\baselineskip}\selectfont \emph{Nan Qi} (qinan@nuaa.edu.cn) received the B.Sc. and Ph.D. degrees from Northwestern Polytechnical University (NPU), China, in 2011 and 2017, respectively. She is currently an assistant professor in Nanjing University of Aeronautics and Astronautics, China. Her research interests include anti-jamming, UAV communications, and relaying networks.\\
\end{spacing}

\begin{spacing}{0.9}
\fontsize{8.5pt}{\baselineskip}\selectfont \emph{Feihuang Chu} (yhb$\_$cfh@nudt.edu.cn) received his M.S. and Ph.D. degree from the Electronic Engineering Institute of PLA, China, in 1997 and 2003, respectively. He is currently working with Space Engineering University, Beijing, China. His research interests include learning theory, and jamming and anti-jamming technology.\\
\end{spacing}

\begin{spacing}{0.9}
\fontsize{8.5pt}{\baselineskip}\selectfont \emph{Shengliang Fang} (mlj$\_$5656@nudt.edu.cn) received his M.S. and Ph.D. degree from Electronic Engineering Institute of PLA, China, in 1998 and 2004, respectively. He is currently working with Space Engineering University, Beijing, China. His research interests include learning theory, and anti-jamming technology.\\
\end{spacing}

\begin{spacing}{0.9}
\fontsize{8.5pt}{\baselineskip}\selectfont \emph{Ximing Wang} (ximingw@nudt.edu.cn) received his Ph.D. degree from Army Engineering University of PLA, China, in 2020. He is currently working with the National University of Defense Technology. His current research interests include anti-jamming, and machine learning.\\
\end{spacing}

\begin{spacing}{0.9}
\fontsize{8.5pt}{\baselineskip}\selectfont \emph{Shuli Ma} (3120160378@bit.edu.cn) received his Ph.D. degree from Beijing Institute of Technology in 2020. She is currently the lecture in Space Engineering University. Her main research interests include image reconstruction, and anti-jamming.\\
\end{spacing}

\begin{spacing}{0.9}
\fontsize{8.5pt}{\baselineskip}\selectfont \emph{Shuo Feng} (fengs13@mcmaster.ca) received his Ph.D. degree from McMaster University, Canada, in 2019. He is currently a Research Scientist with the Nanjing University of Aeronautics and Astronautics. His research interests include cognitive dynamic systems, and machine learning.\\
\end{spacing}


\begin{thebibliography}{1}
%%    ×ÛÊö3-4ƪ
%%%%%%%%%%%%%%%%%%%%%%%%%%%%%%%%%%%%%%%%%%%%%×Ô¼º²Î¿¼ÎÄÏ×%%%%%%%%%%%%%%%%%%%%%%%%%%%%%%%%%%%%%%%%%%%%%%%%%%%%%%%%%%%%%%%%%%%%%%%%%%%%%%%
%%%%%%%%%%%%%%%%%%%%%%%%%%%%%%%%%%%%%%%%%%%%%%%%%%%%%%%%%%%%%%%%%%%%%%
\bibitem{existwork1}
Y. E. Sagduyu, R. A. Berry, and A. Ephremides, ``Jamming games in wireless networks with incomplete information," \emph{IEEE Commun. Mag.}, vol. 49, no. 8, pp. 112-118, Aug. 2011.


\bibitem{existwork2}
K. Pelechrinis, M. Iliofotou, and S. V. Krishnamurthy, ``Denial of service attacks in wireless networks: the case of jammers," \emph{IEEE Commun. Surveys Tutorials}, vol. 13, no. 2, pp. 245-257, Second Quarter, 2011.


\bibitem{existwork3}
K. Grover, A. Lim, and Q. Yang, ``Jamming and anti-jamming techniques in wireless networks: a survey," \emph{Int. J. Ad Hoc and Ubiquitous Comput.}, vol. 17, no. 4, pp. 197-215, Dec. 2014.



\bibitem{existwork4}
M. Aref, S. Jayaweera, \emph{et al.}, ``Survey on cognitive anti-jamming communications," \emph{IET Commun.}, vol. 14, no. 18, pp. 3110-3127, Nov. 2020.



\bibitem{existwork5}
X. Wang, J. Wang, \emph{et al.},``Dynamic spectrum anti-jamming communications: challenges and opportunities," \emph{IEEE Commun. Mag.}, vol. 58, no. 2, pp. 79-85, Feb. 2020.


\bibitem{existwork6}
Y. Xu, Y. Xu, X. Dong, \emph{et al.}, ``Convert harm into benefit: a coordination-learning based dynamic spectrum anti-jamming approach," \emph{IEEE Trans. Veh. Technol.}, vol. 69, no. 11, pp. 13018-13032, Aug. 2020.


\bibitem{existwork7}
Y. Gao, Y. Xiao, M. Wu, \emph{et al.}, ``Game theory-based anti-jamming strategies for frequency hopping wireless communications," \emph{IEEE Trans. Wireless Commun.}, vol. 17, no. 8, pp. 5314-5326, Aug. 2018.



\bibitem{existwork8}
L. Xiao, T. Chen, J. Liu, and H. Dai, ``Anti-jamming transmission stackelberg game with observation errors," \emph{IEEE Commun. Lett.}, vol. 19, no. 6, pp. 949-952, Jun. 2015.



\bibitem{existwork9}
L. Jia, Y. Xu, \emph{et al.},``Stackelberg game approaches for anti-jamming defence in wireless networks," \emph{IEEE Wireless Commun. Mag.}, vol. 25, no. 6, pp. 120-128, Dec. 2018.



\bibitem{existwork10}
L. Jia, Y. Xu, Y. Sun, \emph{et al.},``A game-theoretical learning approach for anti-jamming dynamic spectrum access in dense wireless networks," \emph{IEEE Trans. Veh. Technol.}, vol. 68, no. 2, pp. 1646-1656, Feb. 2019.


\bibitem{existwork11}
F. Yao, and L. Jia,``A collaborative multi-agent reinforcement learning anti-jamming algorithm in wireless networks," \emph{IEEE Commun. Lett.}, vol. 20, no. 10, pp. 1991-1994, Oct. 2016.



%\bibitem{existwork12}
%C. Han, L. Huo, X. Tong, H. Wang, ``Spatial anti-jamming scheme for internet of satellites based on the deep reinforcement learning and Stackelberg game," \emph{IEEE Trans. Veh. Technol.}, vol. 69, no. 5, pp. 5331-5342, May. 2020.


\bibitem{existwork12}
N. Qi, W. Wang, M. Xiao, \emph{et al.}, ``A learning-based spectrum access Stackelberg game: friendly jammer-assisted communication confrontation," \emph{IEEE Trans. Veh. Technol.}, vol. 70, no. 1, pp. 700-713, Jan. 2021.


\bibitem{existwork13}
Y. Xu, J. Wang, Q. Wu, \emph{et al.}, ``A game-theoretic perspective on self-organizing optimization for cognitive small cells," \emph{IEEE Commun. Mag.}, vol. 53, no. 7, pp. 100-108, Jul. 2015.

\bibitem{existwork14}
S. Feng, P. Setoodeh, and S. Haykin, ``Smart home: cognitive interactive people-centric internet of things," \emph{IEEE Commun. Mag.}, vol. 55, no. 2, pp. 34-39, Feb. 2017.


\bibitem{existwork15}
Y. Xu, Q. Wu, L Shen, \emph{et al.}, ``Opportunistic spectrum access with spatial reuse: graphical game and uncoupled learning solutons," \emph{IEEE Trans. Wireless Commun.}, vol. 12, no. 10, pp. 4814-4826, Oct. 2013.
%%%%%%%%%%%%%%%%%%%%%%%%%%%%%%%%%%%%%%%%%%%%%%%%%%%%%%%%%%%%%%%%%%%%%%%%%%%%%%%%%%%%%%%%%%%%%%%%%%%%%%%%%%%%%%%%%%%%%%%%%%%%%%

\end{thebibliography}
\end{document}